# Tungsten doping-induced phase transition in CVD-grown MoS$_2$ bilayers

Ana Senkić[1,‡], Rachael Keneipp[2], Petra Ivatović[3], Namrata Pradeep[2,4], Vincent Meunier[5], Marko Kralj[1], Marija Drndić[2], Nataša Vujičić[1*]

[1] Centre for Advanced Laser Techniques, Institute of Physics, 10 000 Zagreb, Croatia
[2] Department of Physics and Astronomy, David Rittenhouse Laboratory, University of Pennsylvania, Philadelphia, PA, 19104, USA
[3] Physics Department, Faculty of Natural Sciences, 10 000 Zagreb, Croatia
[4] Department of Chemical and Biomolecular Engineering, Philadelphia, PA, 19104, USA
[5] Department of Engineering Science and Mechanics, Penn State University, University Park, PA, USA

*Email: natasav@ifs.hr

## Abstract

Controlling the crystal phase of two-dimensional (2D) transition metal dichalcogenides (TMDs) is essential for tailoring their optical and electronic properties. While phase transitions in monolayer TMDs and semiconductor-to-metal conversions have been widely studied, structural transitions between semiconducting polytypes - particularly in bilayer (2L) systems - remain underexplored. Here, we demonstrate a W doping-induced phase transition from non-centrosymmetric AA stacking to centrosymmetric AB' in 2L MoS$_2$ synthesized by chemical vapor deposition (CVD). Using polarization-resolved second harmonic generation (SHG) and low-frequency Raman spectroscopy, we identify a phase transition correlated with increasing tungsten (W) concentration. The dilute W-doped 2L system exhibits a vanishing SHG signal and a stiffening of the layer-breathing (LB) vibrational mode, in contrast to undoped samples with strong SHG and a softer LB mode. Aberration corrected scanning transmission electron microscopy (AC-STEM) demonstrates the spatial distribution of W concentration and associated structural changes. These findings highlight W-doping as an effective strategy for inducing phase transitions in 2L TMDs, opening new possibilities for engineered heterostructures, phase-controlled device applications or as a source of single photon emitters.

## Introduction

Atomically thin transition metal dichalcogenides (TMDs) have emerged as versatile platforms for exploring novel physical phenomena that arise from reduced dimensionality, strong Coulomb interactions, and sensitivity to external perturbations [1,2]. These 2D materials exhibit tunable optical and electronic properties, making them promising candidates for applications in flexible electronics, spintronics, and quantum technologies [3–5]. Among the various approaches for modulating their properties, chemical doping has attracted significant attention due to its ability to tailor charge carrier densities, influence optical excitations, and even induce structural phase transitions that fundamentally alter the crystal's symmetry and stacking configuration [6–9]. In the extremely low doping limit (< 0.01%), where only one dopant is present in a microscopical area comparable to the diffraction-limited laser spot-size, the dopant acts as a single quantum emitter, as demonstrated in Ref. [10], making such materials promising platforms to study quantum photonics.

Phase transitions in TMDs can also be achieved by carefully tuning synthesis parameters. In chemical vapor deposition (CVD) processes.  For example, the introduction of hydrogen gas [11–13] or a precise control of the growth temperature [6,14] can influence

---

‡ Current affiliation: Institute of Physics, University of Münster, Münster, Germany

growth kinetics and thus enable controllable phase transformations. While extensive studies have been conducted on semiconductor-to-metal transitions [13,15–17], less attention has been paid to transformations between different semiconducting phases. A schematic representation of different semiconducting bilayer stackings is given in Figure 1. They differ in crystal symmetry and interlayer coupling due to variations in interlayer spacing [18]. The centrosymmetric stacking orders are AA' (2H), AB' and A'B, while the non-centrosymmetric are AB (or BA; 3R) and AA [19]. Such structural differences can be probed using various optical techniques such as absorbance spectroscopy, low-frequency Raman spectroscopy, and second harmonic generation (SHG). Transmission electron microscopy (TEM) is also widely used to achieve atomic-scale resolution and identify the material's crystal structure [20]. Notably, authors in Ref. [19] demonstrated that distinct 3R and 2H phases can nucleate atop the same monolayer $WSe_2$ during CVD growth.

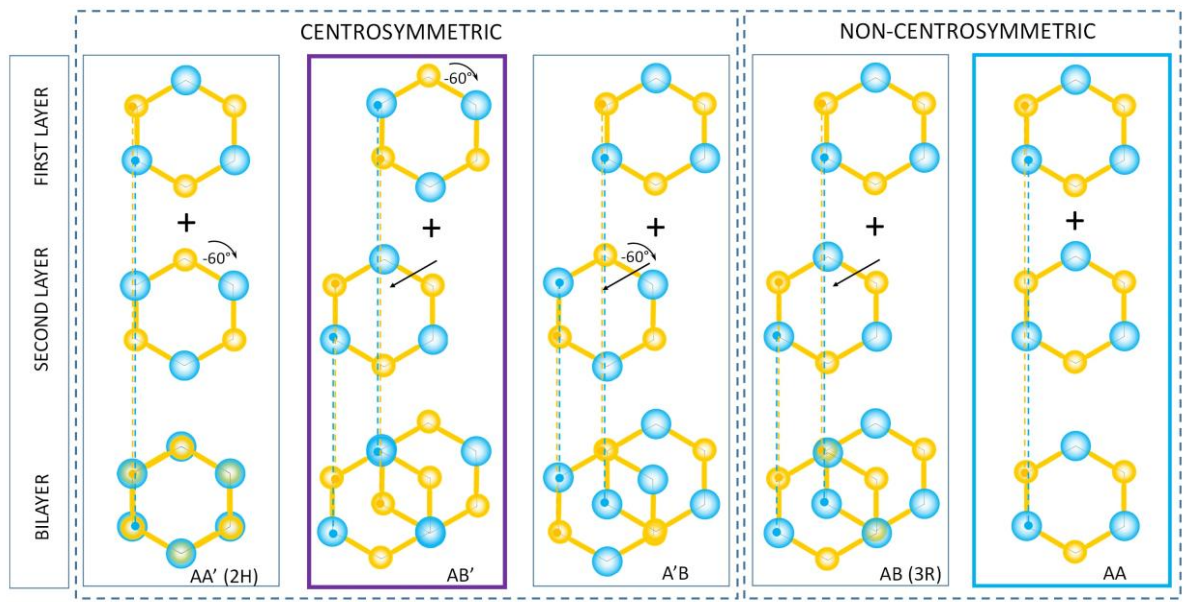

**Figure 1.** Five representative stacking orders in 2L $MoS_2$: AA', AB', A'B, AB and AA. Sulfur (S) is shown in yellow and molybdenum (Mo) in blue. The prime symbol indicates that one layer is rotated by 60 degrees relative to the second one.

In this study, we report a phase transition induced by tungsten (W) doping in 2L CVD-grown $MoS_2$. Undoped samples exhibit a non-centrosymmetric stacking order, confirmed through low-frequency Raman and SHG measurements. Upon introducing W dopants, we observe a phase transition in the 2L structure, evidenced by the stiffening of the layer-breathing (LB) Raman mode and the suppression of the SHG signal. These changes, along with the presence of W dopants and the altered crystal structure, are corroborated by scanning transmission electron microscopy (STEM) analysis. Unlike most prior work that focuses on electronic phase transitions or heavily alloyed systems, our study demonstrates that even low-to-moderate (0.03% - 0.13%) W doping can modulate the crystal phase in 2L $MoS_2$, offering a new pathway for phase control in 2D materials.

## Methods
### Synthesis

Sample synthesis was done using a home-built chemical vapor deposition (CVD) system. Detailed description and instructions for $MoS_2$ growth can be found in our previous work [21]. For the $MoS_2$ growth, only sodium molybdate (Merk's Reagenzien) was used with a 10 ppm concentration. as the molybdenum precursor. To make a liquid-based tungsten precursor, $H_2WO_4$ (99%, Sigma Aldrich) was first diluted in $NH_3$ (1:9). Afterwards, this solution was reduced to a 200 ppm concentration with DI water. After both precursors for molybdenum and tungsten were prepared, two droplets, each having 10 µL, were drop casted on Si/SiO$_2$ wafer, on opposite sides of the substrate. In this way, we obtained two regions: undoped MoS$_2$ and W-doped MoS$_2$. The growth temperature (T$_G$) was set to 850°C. The growth was done under argon flow, with a flow rate during heating to T$_G$ and cooling after growth of 100 sccm, and a flow rate during growth of 75 sccm. Finally, the sulfur temperature during growth time (5 min) was 140°C, and the sulfur heater was turned on during the cooldown.

**Raman and photoluminescence (PL) measurements**

Commercial Renishaw in-via Raman setup was used for Raman and photoluminescence (PL) measurements in a back-scattered configuration. The setup is equipped with a 532 nm (2.33 eV) continuous wave laser source, specialized Bragg filters for low-frequency measurements (down to 15 cm$^{-1}$), and three gratings with 150, 600 and 2400 mm$^{-1}$ constants. For Raman (PL) measurements grating with 2400 (150) mm$^{-1}$ constant was used, with the 100x objective with numerical aperture (NA) 0.9, 600 (460) µW laser power and 0.5 s (0.1 s) acquisition time. All spectra were obtained at room temperature.

**Second harmonic generation (SHG) microscopy**

The nonlinear SHG measurements were conducted with a custom-built scanning confocal optical microscope in a back-scattered geometry. The fundamental beam (1044 nm, 200 fs, 80 MHz) was linearly polarized and focused by a microscopic objective (50x long working distance objective lens; numerical aperture, 0.50) to the spot size of about 1.2 µm in diameter onto the sample at normal incidence. The reflected nonlinear optical signal was collected with the same objective. After passing through a dichroic beam splitter and band-pass filter, the fundamental beam was filtered out, and the nonlinear optical signal was detected by a fiber-coupled spectrograph equipped with a thermoelectrically cooled silicon electron multiplying charge-coupled device (EMCCD). To measure the azimuthal anisotropy pattern of the nonlinear optical signal, we rotated the polarization of the fundamental beam with respect to the sample surface normal using an achromatic half-wave waveplate while the analyzer was kept at the fixed position.

**Scanning transmission electron microscopy (STEM)**

AC-STEM imaging was carried out on a JEOL NEOARM AC-STEM operating at an acceleration voltage of 80 kV and a probe current of ~37 pA. Images were taken with a Gatan annular dark field detector using Gatan's GMS software using a collection angle of 35 – 128 mrad. Dwell times of 8 µs/px or 16 µs/px with 1024 × 1024 pixel image size were used for image collection. All STEM images were processed in ImageJ, and a Gaussian blur filter was applied to all atomic resolution images.

**Density functional theory (DFT) calculations**

Density functional theory (DFT) calculations were performed using the gpaw [22,23] package with analysis tools provided by ase [24]. gpaw is a Python package that utilizes the grid-based projector augmented wave (PAW) method to describe electron–ion interactions [25]. We use the PBE functional to represent the exchange-correlation potential [26], along with the D4 correction to account for van der Waals interaction [27]. For the electronic and phonon calculations, we used a plane-wave basis energy cutoff of 800 eV. We first optimized the lattice parameter of the hexagonal lattice (point group 6/mmm) and found $a$ = 3.154 Å, while the dimension of the cell perpendicular to the membrane was kept at $c$ = 33 Å to avoid spurious effects arising from periodic boundary conditions. For primitive cell calculations, the Brillouin zone (BZ) was sampled with a 9x9x1 Monkhorst–Pack grid, which was found to be sufficient for convergence [28].

For the calculations of the quasi-harmonic phonon band structure, we used the package phonopy [29] to calculate and process the force constants obtained using the finite difference method (FDM) with the gpaw calculator. Zone-centered phonons were calculated using a 3x3x1 supercell. In this case, a 3x3x1 $k$-point grid was used. A displacement distance of 0.01 Å was adopted for the calculation of the force constants within the FDM. In a post-processing step, we used the code hiPhive [30] with the maximum estimated cluster cutoff to enforce the rotational sum rules on the force constants. To minimize numerical errors and ensure minimal drift forces, strict convergence criteria of $10^{-8}$ were used for both the wavefunction and density convergence. To calculate the phonons in the doped cases, we used a 4x4x1 supercell with on W substituted for on Mo.

## Results

Figure 2 presents PL spectra of undoped and W-doped 2L $MoS_2$ samples. Optical micrographs of these samples are shown in Figure S1 in Supplementary Information (SI). Typical intralayer A and B excitons are present at $(1.8184 \pm 2 \cdot 10^{-4})$ eV and $(1.971 \pm 0.001)$ eV, respectively, for the undoped 2L $MoS_2$. In the case of the W-doped sample, the A and B exciton energies are $(1.8353 \pm 3 \cdot 10^{-4})$ eV and $(2.004 \pm 0.002)$ eV. These intralayer excitons are located at the band edge and present direct transitions from the conduction band minimum (CBM) to the valence band maximum (VBM) [31]. The energy difference between A and B excitons in the undoped $MoS_2$ sample is 152.6 meV and in W-doped $MoS_2$ is 168.7 meV. This difference in the energies of excitonic emission lines is a consequence of the different phase structures in the two materials, given that the centrosymmetric phase has stronger spin–orbit coupling [32]. Additionally, in the W-doped $MoS_2$ sample, the A exciton intensity is reduced, while its full width at half maximum (FWHM) is increased, indicating the existence of additional dopant-induced energy levels even at room temperature [33,34].

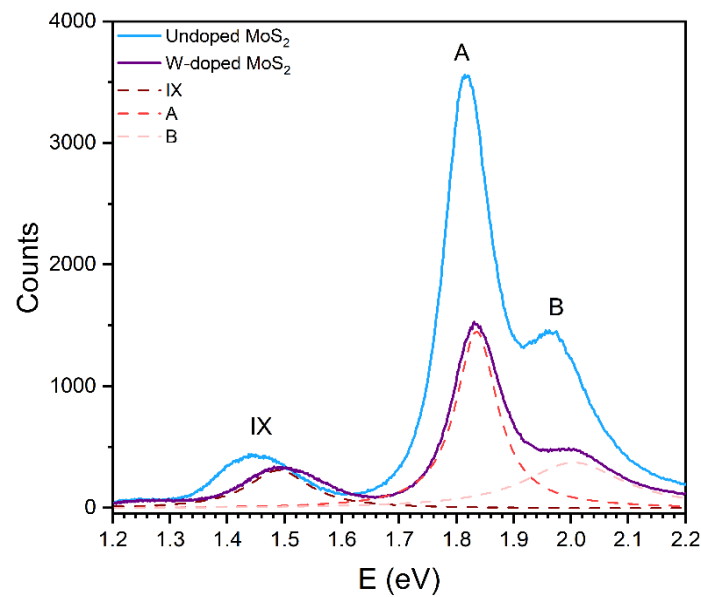

**Figure 2.** Photoluminescence (PL) spectra of undoped (blue line) and W-doped (purple line) 2L $MoS_2$. Fit components of the W-doped spectra are shown in dashed lines.

In 2L samples additional emission appears, which corresponds to an indirect transition [35]. This emission line, called the interlayer exciton (IX) is blueshifted by 48 meV in the presence of W atoms. This shift is significantly larger than the blueshifts of the A and B excitons. Contrary to the behavior of intralayer excitons, the IX is highly sensitive to the interlayer coupling [36,37] and hence provides insights into structural changes. Prior studies [38–40] have shown that the IX exciton in homobilayers exhibits a blueshift in the 2H polymorph and a redshift in the 3R configuration, attributed to electronic band hybridization effects.

To determine the influence of W-doping on the vibrational spectra in 2L $MoS_2$, we further analyzed Raman spectra. In the high-frequency part (> 100 $cm^{-1}$), 2L $MoS_2$ exhibits two modes: $E_{2g}$ and $A_{1g}$, which correspond to in-plane and out-of-plane intralayer oscillations of molybdenum and sulfur atoms, respectively [41–43]. Figure S2 in the SI shows Raman spectra of undoped and W-doped 2L $MoS_2$ samples. Spectra were fitted using sum of two Lorentzian functions, yielding $E_{2g}$ and $A_{1g}$ mode frequencies of $(384.16 \pm 0.03)$ $cm^{-1}$ and $(405.95 \pm 0.04)$

cm$^{-1}$ for undoped 2L MoS$_2$, and (383.42 ± 0.05) cm$^{-1}$ and (404.51 ± 0.08) cm$^{-1}$ for W-doped 2L MoS$_2$. Frequencies of both modes soften - indicative of presence of strain and (chemical) doping due to W atoms, respectively [44]. On the other hand, the FWHM of A$_{1g}$ mode is increased in case of W-doped 2L MoS$_2$, consistent with previous reports on doping [45].

Low-frequency modes (< 100 cm$^{-1}$) are ascribed to collective oscillations of the individual layers in multilayer systems [46,47]. The frequencies of these modes depend strongly on the number of layers [48] and are further influenced by crystal symmetry and twist angle [49,50], meaning that this frequency range can be used as fingerprint for determining the crystal symmetry and interlayer coupling. The lower-frequency mode, called the shear (S) mode, involves lateral oscillations between layers, whereas the higher-frequency layer-breathing (LB) mode corresponds to out-of-plane layer oscillations.

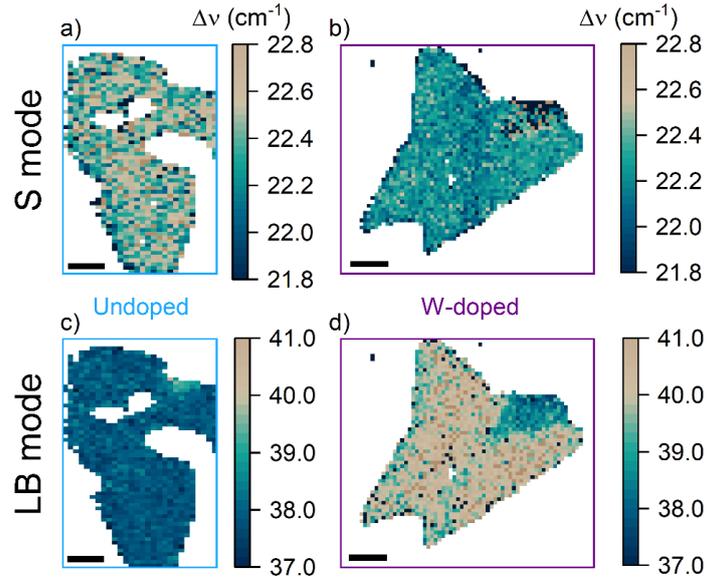

**Figure 3.** Raman maps of interlayer modes' frequency (cm$^{-1}$) are shown for undoped (a,c) and W-doped (b,d) 2L MoS$_2$. Scalebar is 3 μm.

Figure 3 presents spatial maps of S and LB mode frequencies for the two 2L samples: undoped MoS$_2$ (Fig. 3a, c) and W-doped MoS$_2$ (Fig. 3b, d). While the S mode frequency remains mainly unchanged, the LB mode frequency increases by approximately 2.5 cm$^{-1}$ in the W-doped 2L MoS$_2$, from 38 to 40.5 cm$^{-1}$. Previous studies have reported that 2L MoS$_2$ with 2H (3R) phase exhibits LB mode frequencies above (below) 40 cm$^{-1}$ [46,49,51–54] . This raises the possibility that W doping causes the phase transition in CVD-grown 2L MoS$_2$. Since both samples exhibit significant PL emission, we exclude the existence of 1T metallic phase.

To determine if the LB frequency can be continuously tuned with the W concentration, we measured Raman maps on several samples throughout the substrate (Figure S3 in the SI). Most islands on the side of the substrate where only W precursor was dropcasted exhibit elevated LB mode frequencies, suggesting the presence and influence of W dopants. Intermediate LB frequencies are observed in the central region, whereas the undoped region, where only Mo precursor was dropcasted, displays the lowest LB mode frequencies. Intermediate LB frequencies could indicate the presence of both polytypes within the two layers, which can occur under conventional CVD growth conditions [19]. DFT calculations at the PBE-D4 level of theory for the various stackings are summarized in Table S1. The results of DFT calculations indicate that the LB mode blueshifts by about 2.88 cm$^{-1}$ from AA in the undoped 2L MoS$_2$ to AB' in the W-doped sample. The corresponding calculated unfolded band structures for W-doped 2L MoS$_2$ are shown in Figure S4.

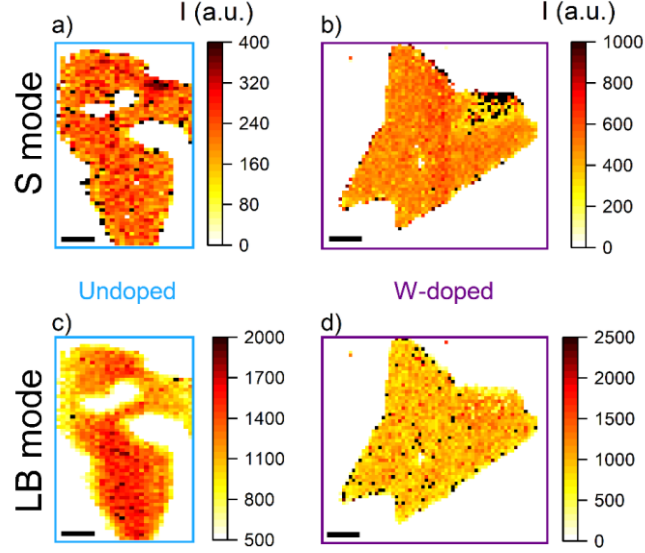

**Figure 4.** Raman maps of interlayer modes' (integrated) intensity are shown for undoped (a,c) and W-doped (b,d) 2L MoS$_2$. Scalebar is 3 μm.

Apart from the change in the LB frequency, a noticeable change in the intensity of the S mode is also observed (Figure 4). Namely, in the W-doped region, the intensity of S mode is several times larger than in the undoped region, suggesting that the interlayer distance is smaller. This means that the layer-layer vibrations cause a larger polarizability change, resulting in the higher intensity [46]. This increase in the S mode intensity can also be correlated with the larger IX energy (Figure 2). Furthermore, DFT calculations of the van der Waals energy per unit cell and the corresponding equilibrium interlayer distances $h$ for different stacking orders in W-doped 2L MoS$_2$ confirm that the interlayer spacing is smaller in AB' stacking than in AA stacking (Table 1). We would also like to note that the relative changes between different stackings are similar to the undoped 2L MoS$_2$. This reduction in spacing distance corroborates the preceding discussion.

**Table 1.** The van der Waals energy per bilayer unit cell and corresponding interlayer distance $h$ at equilibrium for each stacking order.

| Stacking order | AA' | AA | A'B | AB | AB' |
| --- | --- | --- | --- | --- | --- |
| E (eV/unit-cell) | -0.22 | -0.13 | -0.22 | -0.14 | -0.20 |
| h (Å) | 6.132 | **6.709** | 6.081 | 6.969 | **6.169** |

To verify the change in crystal symmetry, we performed polarization-dependent SHG measurements (Figure 5). SHG is a nonlinear optical process that is highly sensitive to crystal symmetry, as it is allowed only in non-centrosymmetric materials under the electric-dipole approximation [55,56]. Since AA, AB (3R phase) and BA stacking are non-centrosymmetric, they consistently show a strong SHG signal [57]. In contrast, AA' (2H phase), AB' and A'B stackings are centrosymmetric for even-numbered layers, while they are non-centrosymmetric for odd-numbered layers larger than one. This means that for 2L samples, only non-centrosymmetric MoS$_2$ will show SHG signal. One additional difference between most studied crystal phases, 3R and 2H, is that the SHG intensity in 3R phase increases with the increase of layer number (I(1L) < I(2L) < I(3L)), while in 2H phase it decreases as the number of (odd) layers increases (I(1L) > I(3L) > I(5L)) [57].

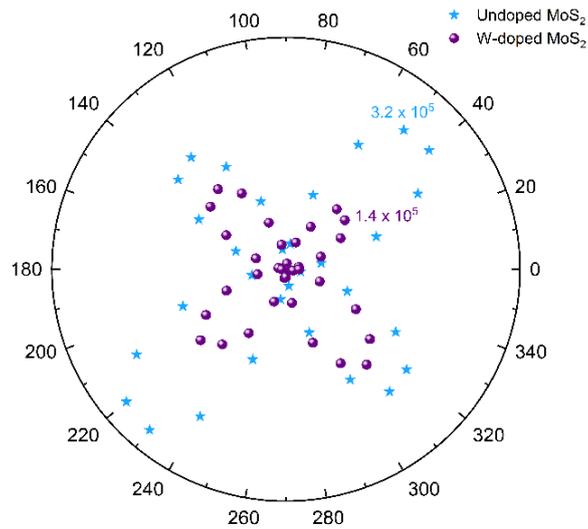

**Figure 5.** Polar plot of the local SHG intensity with respect to the laser polarization. The laser power for the W-doped sample was increased 2.5 times.

For our samples, under identical acquisition conditions, the W-doped 2L $MoS_2$ showed no detectable SHG signal, in contrast to the undoped 2L. This is in agreement with results from low-frequency Raman spectra that the W-doping induces a phase transition in 2L $MoS_2$ from non-centrosymmetric to centrosymmetric and changes its interlayer coupling. It is worth noting that a weak SHG response in W-doped 2L sample could only be observed after increasing the laser power by a factor of 2.5, similar to report [19] where authors needed to increase the exposure time in order to observe SHG signal in 2H samples. Polarization-resolved SHG measurements (Figure 5) reveal a four-fold rotational symmetry in both doped and undoped samples, consistent with the in-plane crystal symmetry of $MoS_2$ in the back-scattered geometry and the experimental realization of polarization measurements.

To quantify the spatial distribution of W dopants across the substrate, we performed STEM measurements. Figure S5 in the SI shows STEM images of 1L and 2L $MoS_2$ with varying W concentrations depending on their position on the substrate. In the W-doped region, the atomic percentage of W is measured to be about 0.13%, while in the intermediate region, located centrally on the substrate, the W concentration is as low as 0.03%. Detailed description on how these images were analyzed can be found in the SI. As expected, no significant presence of W atoms is observed in the undoped region. However, in isolated cases - such as in Figure S5 (b) in the SI - individual W atoms can still be detected within small 10 × 10 nm areas. Materials with such ultra-low dopant concentrations are of particular interest for quantum photonics applications, as they may host single-photon emitters, as recently reported in [10].

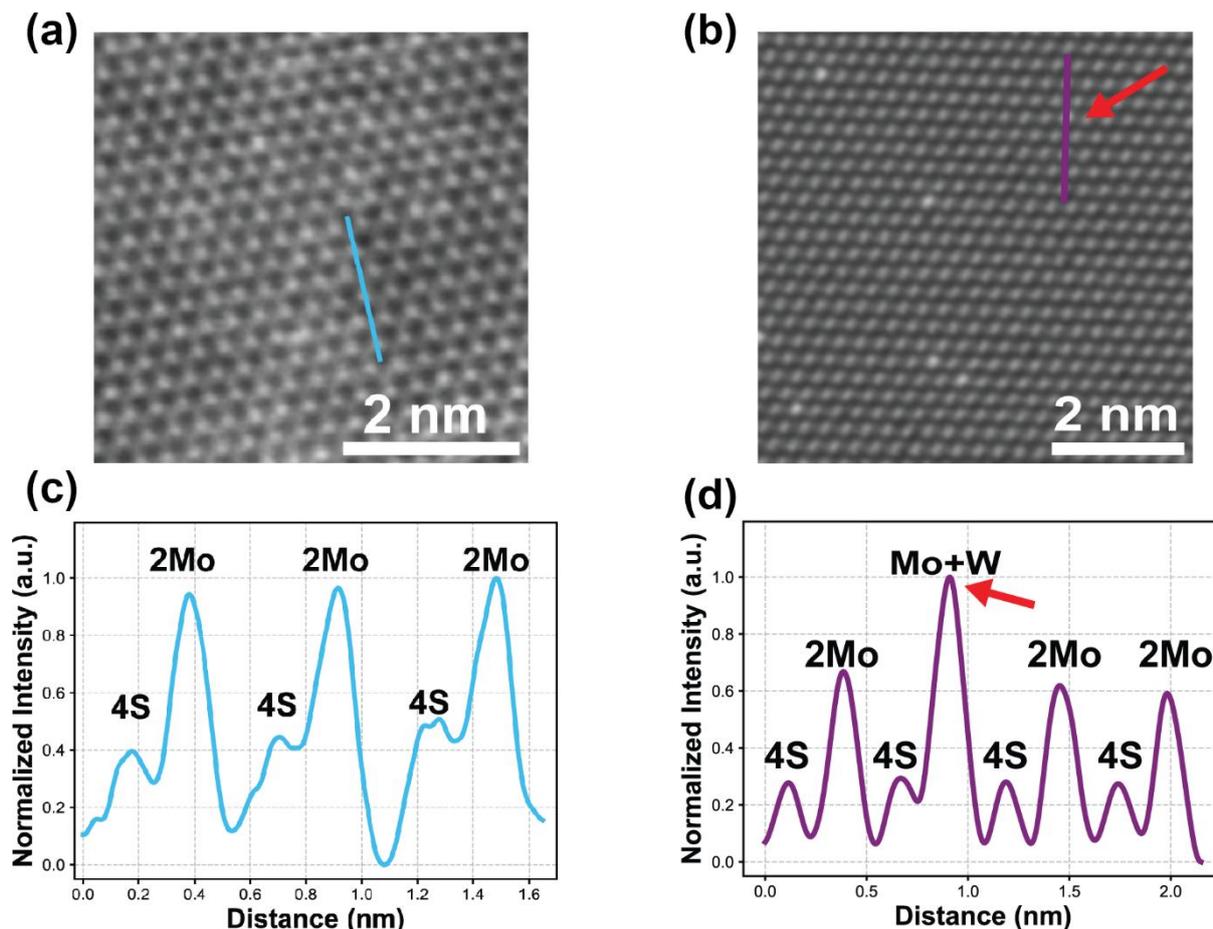

**Figure 6.** a) Atomic resolution STEM image of 2L MoS$_2$ showing the AA stacking with inset blue intensity line profile that is plotted in (c). b) AC-STEM images of W-doped 2L MoS$_2$ showing the AB' stacking with inset purple intensity line profile that is plotted in (d). c) Intensity profile of the blue line profile inset in (a) showing the intensity peaks due to the overlapping Mo and S atoms. d) Intensity profile of the purple line profile inset in (b) showing the increase in ADF intensity due to the presence of W-dopants.

To further support the hypothesis of a doping-induced phase transition, we also conducted STEM imaging on 2L samples from the same substrate regions (Figure 6). In the undoped region, 2L MoS$_2$ samples dominantly have non-centrosymmetric AA stacking, but as shown in Figure S6 some of them also exhibit 2H phase (AA' stacking), without evident presence of W atoms. Such samples with different phases were also observed with spatial Raman maps (Figure S7). Even though the AA stacking is not the most stable one, it is possible that the presence of sodium in the molybdenum precursor lowers the formation energy of this stacking order, making it more favorable than the AA' or AB. In the most extreme case, tuning the concentration of alkali elements in the metallic precursor causes the transition from 2H to 1T' phase MoS$_2$ and WS$_2$ monolayers, as shown in [13]. Furthermore, a recent paper has shown that Na adsorption within the interlayers of a bulk 2H-MoS$_2$ can alter the arrangement of adjacent layers, facilitating an efficient phase transition from the 2H phase to the energetically less favorable 3R phase [58].

In the W-doped region, 2L samples do not exhibit typical 2H phase, but modified centrosymmetric structure which is labeled as AB' [19]. Recent theoretical models [59] show that interstitial Mo impurities change the stacking order in 2L MoS$_2$. The main result is that the Mo impurity strongly hybridizes with the nearest atoms, thus changing the most stable stacking configuration from AA' (2H) to trigonal AB'. Since we have W-doping it is possible that significantly larger atom or its different level hybridization in a crystal lattice influence the change from non-centrosymmetric to symmetric structure.

These findings suggest that within a single substrate, 2L MoS$_2$ samples can exhibit different phases, governed by the local W concentration. In W-doped region the 2L have centrosymmetric AB' stacking. This is supported by the absence of an SHG signal and the observed increase in the LB Raman mode frequency. In the central part of the substrate, again, depending on the local W concentration, mixed phases in 2L samples can occur, consistent with the intermediate LB frequency and moderate W concentration (Figure S5 b). In contrast, 2L in the undoped region (a part of the substrate with only Mo precursor) have strong SHG due to the broken inversion symmetry and a softened LB mode, indicating a non-centrosymmetric stacking order.

In our CVD growth procedure, which includes Na as a promoter [21], undoped $MoS_2$ consistently crystallize in the non-centrosymmetric phase, as confirmed by SHG measurements and as was previously reported in [60]. The centrosymmetric phase appears only upon W doping, suggesting that W incorporation plays a direct role in stabilizing this stacking order. Previous studies on $Mo_{1-x}W_xS_2$ alloys [61–64] report phase coexistence and transitions depending on the W content (particularly around $x \approx 0.5$), as well as on growth temperature and synthesis protocols. A doping-induced structural transition from 2H to 3R phase in layered $MoS_2$, accompanied by a renormalization of the valence band structure due to Nb dopants - from monolayer to bulk - has been reported in [65]. Analogous behavior has been reported in $Fe_2P$ nanorods, where Mo doping induces a structural transition from hexagonal to orthorhombic symmetry. Specifically, mixed-phase regions are observed for Mo concentrations in the range x = 0.03–0.09, with a complete transformation to the orthorhombic phase occurring at $x \approx 0.11$ [66]. This transition is attributed to strain induced by dopant incorporation.

## Conclusion

In conclusion, we observed that dilute W-doping (0.03% - 0.13%) in 2L $MoS_2$ causes a structural phase transition from non-centrosymmetric AA stacking to centrosymmetric AB', confirmed with second harmonic generation (SHG), which revealed a significant decrease in SHG intensity consistent with the presence of inversion symmetry in centrosymmetric 2L. Complementary low-energy Raman measurements further support this phase transition by detecting changes in the interlayer vibrational modes, particularly in the frequency shift of the LB mode and in the intensity of the S mode.

The concentration of W atoms across the entire substrate was determined from the atomically resolved STEM images and correlated with the frequency of LB mode, suggesting that even low levels of W doping can significantly influence the stacking order and interlayer coupling in 2L $MoS_2$. These findings underscore the sensitivity of $MoS_2$ 2L stacking configurations to atomic-scale perturbations and highlight the utility of W-doping as a means of phase engineering in the host material. Our further work will involve investigation of the interlayer coupling strength with temperature-dependent optical measurements.

## Acknowledgements

This research was led by DOE grant DE-SC0023224 for investigating electron microscopy; M.D. and R.K. acknowledge primary support from this grant. Electron microscopy was performed in the Singh Center for Nanotechnology, which houses shared facilities supported by the NSF National Nanotechnology Coordinated Infrastructure Program under grant NNCI-2025608, and the electron microscopy facility is partially supported by the University of Pennsylvania Materials Research Science and Engineering Center (MRSEC) DMR-2309043.

M.D. and R.K gratefully acknowledge Prof. Eric Stach for discussions of AC-STEM data. R.K. acknowledges the graduate fellowship from the Vagelos Institute for Energy Science and Technology at the University of Pennsylvania.

The Croatian team was supported by projects CALT, (grant no. KK.01.1.1.05.0001) and CEMS (grant no. PK.1.1.10.0002) co-financed by the European Union through the European Regional Development Fund—Competitiveness and Cohesion Operational Program and Croatian Ministry of Science, Education and Youth. A. S., N. V. and M. K. acknowledge financial support from the project "Podizanje znanstvene izvrsnosti Centra za napredne laserske tehnike (CALTboost)," financed by the European Union through the National Recovery and Resilience Plan 2021–2026 (NRPP), and IP- 2022-10-4724 supported by the Croatian Science Foundation. A.S., P.I. and N.V. gratefully acknowledge Dr. Dario Novoselović for providing equipment (a femtosecond oscillator for the SHG setup) through a collaboration agreement between the Agricultural Institute in Osijek and the Institute of Physics. A.S. and N.V. gratefully acknowledge the support from Borna Radatović.

## Author contribution

Ana Senkić (orcid.org/0000-0002-0567-5299): conceptualization, methodology, investigation and data analysis - synthesis, optical measurements, systematization, visualization, writing and editing; Rachael Keneipp (orcid.org/0009-0008-7785-8714): conceptualization, methodology, investigation and data analysis: STEM, visualization, writing and editing; Petra Ivatović (orcid.org/0009-0006-3749-7024) : experimental realization (building SHG setup), editing; Namrata Pradeep (orcid.org/0009-0003-7150-5778): data analysis: STEM, writing and editing; Vincent Meunier (orcid.org/0000-0002-7013-179X): conceptualization, editing; Marko Kralj (orcid.org/0000-0002-9786-3130): financial acquisition, writing and editing; Marija Drndić (orcid.org/0000-0002-8104-2231): supervision, methodology, conceptualization, writing and editing; Nataša Vujičić (orcid.org/0000-0002-5437-5786): supervision, methodology, conceptualization, writing and editing.

# SUPPLEMENTARY INFORMATION

# Tungsten doping-induced phase transition in CVD-grown MoS$_2$ bilayers

Ana Senkić[1,‡], Rachael Keneipp[2], Petra Ivatović[3], Namrata Pradeep[2,4], Vincent Meunier[5], Marko Kralj[1], Marija Drndić[2], Nataša Vujičić[1*]

[1] Centre for Advanced Laser Techniques, Institute of Physics, 10 000 Zagreb, Croatia
[2] Department of Physics and Astronomy, University of Pennsylvania, Philadelphia, PA, 19104, USA
[3] Physics Department, Faculty of Natural Sciences, 10 000 Zagreb, Croatia
[4] Department of Chemical and Biomolecular Engineering, Philadelphia, PA, 19104, USA
[5] Department of Engineering Science and Mechanics, Penn State University, University Park, PA, USA

*Email: natasav@ifs.hr

## Additional figures

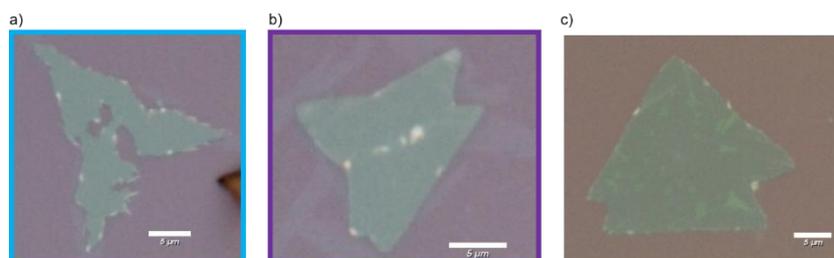

**Figure S1.** Optical micrographs of a) undoped and b) W-doped bilayer MoS$_2$ samples. Optical contrast reference for monolayer sample with isolated bilayer islands is given in c). Scale bars are 5 μm.

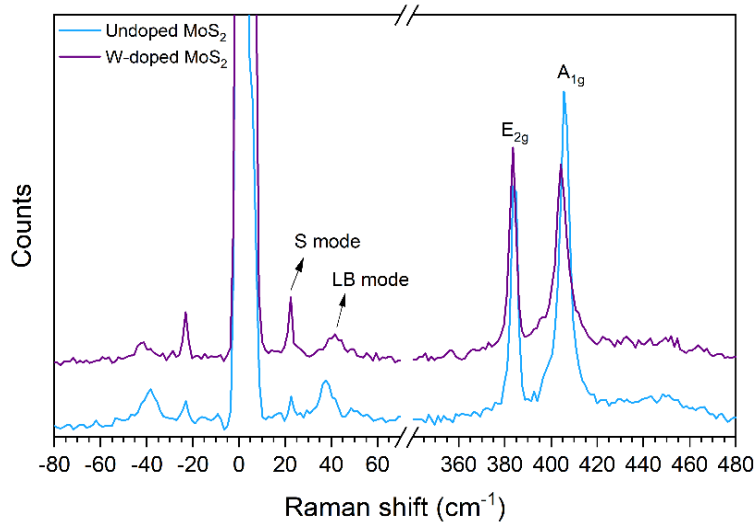

**Figure S2.** Room temperature Raman spectra of undoped and W-doped MoS$_2$ bilayers.

---

‡ Current affiliation: Institute of Physics, University of Münster, Münster, Germany

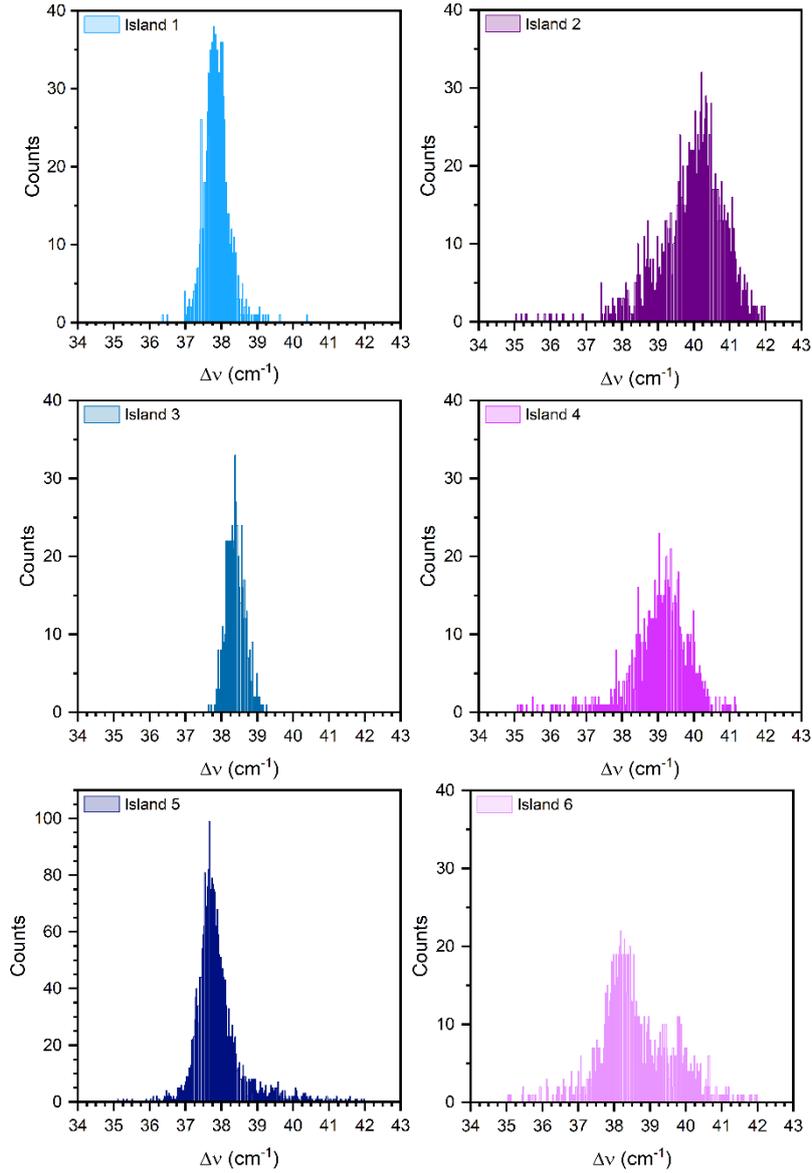

**Figure S3.** Histogram of frequencies of layer-breathing (LB) mode in $MoS_2$ bilayers with different concentrations of tungsten doping. Islands 1 and 2 correspond to undoped and W-doped $MoS_2$ bilayer samples in the main text, respectively.

**Table S1.** Frequencies ($cm^{-1}$) of all Raman active modes as calculated by DFT in undoped and in W-doped 2L $MoS_2$ in various high-symmetry stacking. The W-doped system corresponds to 3.1% doping.

|  | Undoped 2L $MoS_2$ | | | | W-doped 2L $MoS_2$ | | | | |
|---|---|---|---|---|---|---|---|---|---|
|  | Shear | Layer-breathing | $E_{1g}$ | $A_{2g}$ | Shear | Layer-breathing | $E_{1g}$ | $A_{2g}$ | |
| AA | - | **33.1** | **380.3** | **403.8** | - | 32.7 | 382.6 | 403.3 | AA |
| AA' | 22.5 | 38.0 | 380.8 | 404.3 | 22.7 | 38.9 | 383.4 | 403.4 | AA' |
| AB | 23.6 | 37.1 | 380.5 | 404.8 | 23.5 | 38.9 | 383.2 | 403.4 | AB |
| A'B | - | 32.3 | 380.8 | 403.8 | - | 31.8 | 382.8 | 403.2 | A'B |
| AB' | 17.8 | 33.0 | 381.0 | 404.5 | **16.5** | **36.0** | **382.9** | **403.5** | AB' |

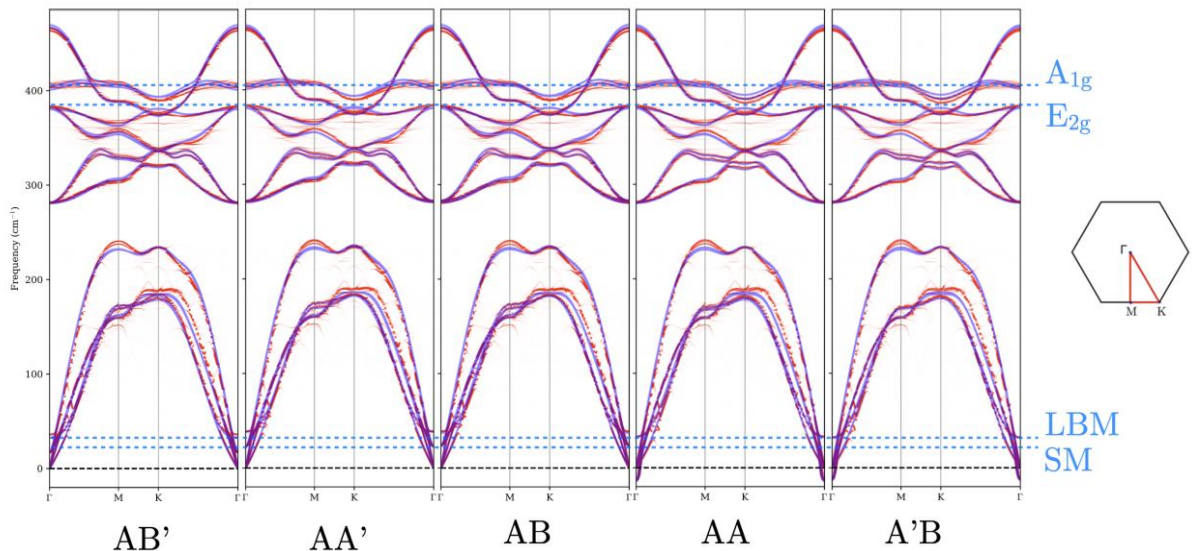

**Figure S4.** Calculated unfolded band structures for W-doped MoS$_2$ bilayer (red) and corresponding bands for undoped (blue) bilayers in various stacking orders.

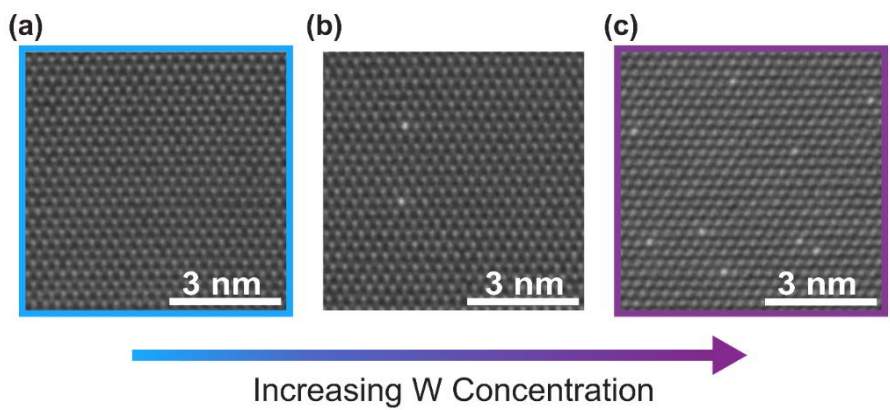

**Figure S5.** Different tungsten concentrations. (a) shows an undoped MoS$_2$ monolayer with no tungsten present. (b) shows two W atoms present within a monolayer of the MoS$_2$ and the W concentration increases in the bilayer shown in (c).

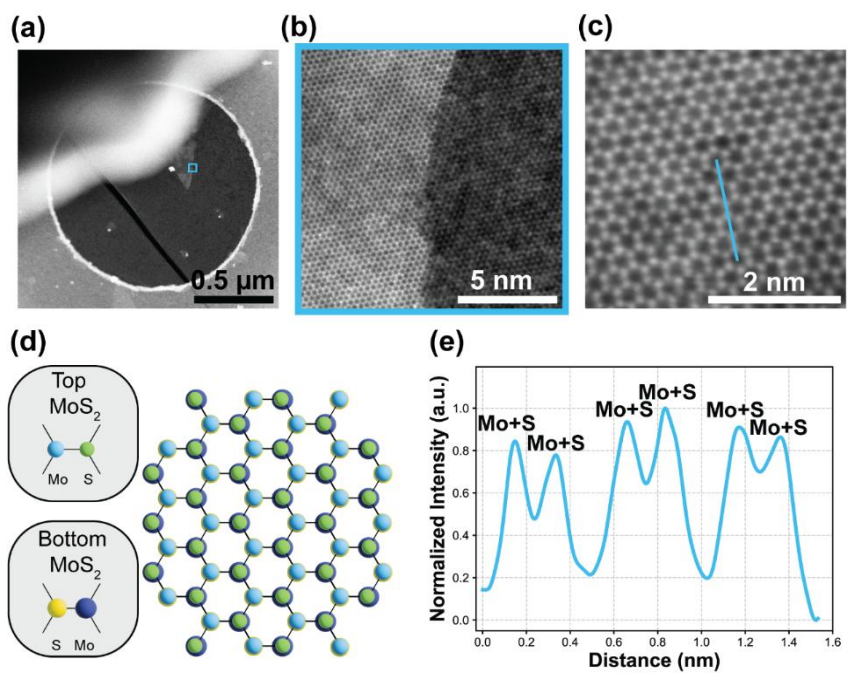

**Figure S6.** AC-STEM images of bilayer AA' MoS$_2$. (a) STEM image of bilayer flake on monolayer MoS$_2$ with inset showing region imaged in (b). (b) Lateral interface of mono- and bi-layer MoS$_2$ regions. (c) Atomic resolution STEM image of MoS$_2$ showing the AA' stacking with inset blue intensity line profile that is plotted in (e). (d) Schematic showing the AA' structure of the MoS$_2$ imaged in (c). (e) Intensity profile of the blue profile inset in (c) showing the intensity peaks due to the overlapping Mo and S atoms.

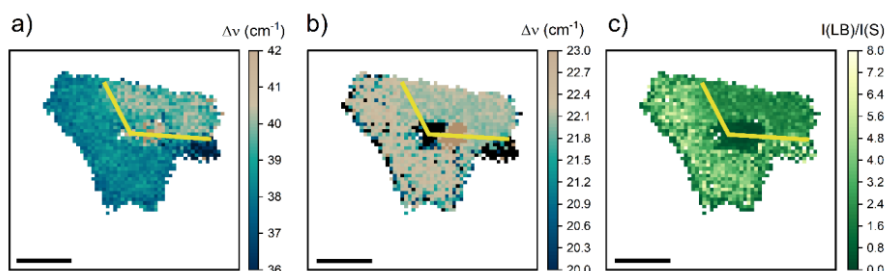

**Figure S7.** Spatial Raman maps of one bilayer MoS$_2$ sample from undoped region, containing both non-centrosymmetric AA and centrosymmetric AB' stacking order. a) Frequency of layer-breathing (LB) mode, b) frequency of shear (S) mode and c) intensity ratio I(LB)/I(S). The increase in LB mode's frequency is accompanied by a drop in its intensity (or intensity increase of shear mode). Middle region is multilayer sample nucleation. Scalebars are 5μm.

## Description of the algorithm for W-concentration determination

To determine the tungsten doping concentration, we first processed the annular dark field (ADF) scanning transmission electron microscope (STEM) images. Each image was pre-processed by applying a Gaussian blur (σ = 2.5 pix) and normalizing the pixel intensities to a range of 0 to 1. Atomic sites were identified using the *blob_log* method from the scikit-image (v0.25) library, which applies a Laplacian of Gaussian (LoG) filter across multiple scales. This approach detects bright atomic columns as local maxima within a three-dimensional LoG response space. Although computationally intensive, this method provides high accuracy in locating atom positions, particularly for small-scale features. Atoms at the image boundaries were excluded.

To distinguish between tungsten (W) and molybdenum (Mo) atoms, several thresholding approaches—such as Otsu, Yen, minimum, and mean methods—were evaluated. Ultimately, a z-score–based classification was found to be the most effective, with atomic sites exhibiting z-scores greater than 3.0 classified as tungsten. To minimize misclassification caused by anomalously bright Mo regions, a spatial correction was applied: atomic sites within five atomic radii of one another were cross-checked to adjust for local intensity contamination. Images were exported with markers at each atomic sight, as shown in Figure S1. The Mo and W atoms identified from this analysis were used to determine the atomic percent of W present (Table S1) in different regions of the sample.

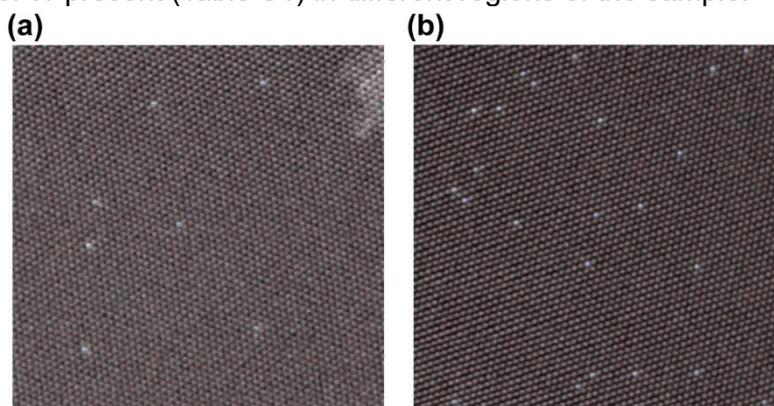

**Figure S8.** AC-STEM images of W) identification within the MoS$_2$ lattice in (a) monolayer and (b) bilayer samples. The red overlays indicate Mo atoms and the blue overlays indicate W atoms.

**Table S2.** Calculation of percent W for different sample regions.

| Sample Region | Measured # of Mo Sites | # of W atoms | Atomic % W | Ratio Mo:W |
|---|---|---|---|---|
| YA Monolayer - 1 | 5190 | 20 | 0.13 | 261:1 |
| YA Monolayer - 2 | 6250 | 24 | 0.13 | 261:1 |
| YA Monolayer - 3 | 9566 | 21 | 0.07 | 457:1 |
| YB Bilayer - 1 | 4492 | 4 | 0.03 | 1123:1 |
| YB Bilayer - 2 | 3119 | 8 | 0.09 | 390:1 |